\documentclass[letterpaper,preprintnumbers,prd,twocolumn,nofootinbib,nobibnotes,showpacs]{revtex4}
\usepackage{amsfonts}
\usepackage{mathrsfs}
\usepackage{epsfig}
\usepackage{graphicx}%
\usepackage{dcolumn}
\usepackage{amsmath}

\makeatletter
\def\btt#1{\texttt{\@backslashchar#1}}%
\DeclareRobustCommand\bblash{\btt{\@backslashchar}}%
\makeatother
\begin{document}

\title{Introduction of the generalized Lorenz gauge condition into the vector tensor theory}
\author{Changjun Gao}\email{gaocj@bao.ac.cn}\affiliation{$^{}$The
National Astronomical Observatories, Chinese Academy of Sciences,
Beijing 100012, China} \affiliation{{$^{}$Kavli Institute for
Theoretical Physics China, CAS, Beijing 100190, China }}

\date{\today}

\begin{abstract}
We introduce the generalized Lorentz gauge condition in the theory
of quantum electrodynamics into the general vector-tensor theories
of gravity. Then we explore the cosmic evolution and the static,
spherically symmetric solution of the four dimensional vector
field with the generalized Lorenz gauge. We find that, if the
vector field is minimally coupled to the gravitation, it behaves
as the cosmological constant. On the other hand, if it is
nonminimally coupled to the gravitation, the vector field could
behave as vast matters in the background of the spatially flat
Friedmann-Robertson-Walker Universe. But it may not be the case.
The weak, strong and dominant energy conditions, the stability
analysis of classical and quantum aspects would put constraints on
the parameters and so the equation of state of matters would be
greatly constrained.

\end{abstract}


\maketitle

\section{Introduction}
The vector-tensor theories of gravity are first proposed by Will,
Nordtvedt and Hellings \cite{will:72}. Then they are used in
cosmology to model inflaton \cite{ford:89}, dark matter
\cite{bek:04} and dark energy \cite{bel:08}. However, it is
recently found that the Lorentz invariant vector-tensor theories
are usually plagued by instabilities \cite{bel:09}
\cite{picon:09}. On the contrary, the Lorentz violation vector
field models \cite{kos:89} are special such that some of these
models are free of instabilities \cite{el:06}. The well-known
Einstein-aether theory \cite {ted:00} belongs to these special
models. The remarkable difference of the Einstein-aether theory
from the usual vector-tensor theory comes from the fact that the
vector field is constrained to have constant norm. This constraint
eliminates a wrong-sign kinetic term for the length-stretching
mode \cite{ell:05}, hence giving the theory a chance to be viable.

The fixed-norm constraint in the Einstein-aether is achieved by
the presence of a Lagrange multiplier field in the Lagrangian
density. The Lagrange multiplier method presents the constraints
on the motion of some physical quantities in nature. So taking
into account some physical constraints on the motion of physical
quantities, one can always introduce the Lagrange multiplier into
the corresponding Lagrange function. Actually, using this method,
Mukhanov, Brandenberger and Sornborger have early proposed a
nonsingular universe by limiting the spacetime curvature to some
finite values \cite{muk:92}.

Except for the fixed-norm constraint, are there any other
constraints on the four-vector (four dimensional vector) field?
The answer is yes. We remember that there is a unified formulation
of quantum electrodynamics which has the Lagrangian as follows
\cite{lau:67}
\begin{eqnarray}
\label{eq:intro}
\mathscr{L}&=&-\frac{1}{4}F_{\mu\nu}F^{\mu\nu}+\lambda\left(\nabla_{\mu}A^{\mu}
-\frac{1}{2}\gamma\lambda\right)\nonumber\\&&-j_{\mu}A^{\mu}+\bar{\psi}\left(i\gamma_{\mu}\partial^{\mu}-m\right)\psi\;.
\end{eqnarray}
Here $j_{\mu}=e\bar{\psi}\partial_{\mu}\psi$ and $F_{\mu\nu}$ is
the field strength tensor of Maxwell field. $\lambda$ is the
Lagrange multiplier field which has the dimension of inverse
length, $l^{-1}$. $\gamma$ is a dimensionless constant. The
$\lambda$ term is the generalized Lorentz gauge. $\gamma=0,\ 1,\
3$ correspond to the Landau gauge \cite{lan:56}, the Feynman gauge
\cite{fey:49} and the Yennie-Fried gauge \cite{fri:58},
respectively. We see the generalized Lorentz gauge condition could
also be understood as a constraint on the divergence of
four-vector field $A^{\mu}$. Motivated by this point, we introduce
the generalized Lorentz gauge condition into the general
vector-tensor theories.

The paper is organized as follows. In section II, we explore the
cosmic evolution and the static, spherically symmetric solution of
the four-vector field which is minimally coupled to the
gravitation. We find the field behaves as a cosmological constant.
In section III, we investigate the cosmic evolution of the
four-vector which is nonminimally coupled to the gravitation and
find it could play the role of vast matters for some appropriate
parameters. Section IV gives the conclusion and discussion.

We shall use the system of units in which $16\pi
G=c=\hbar=4\pi\varepsilon_0=1$ and the metric signature $(-,\ +,\
+,\ +)$ throughout the paper.
\section{Cosmological Constant from the four-vector field with the Generalized Lorenz Gauge}
\subsection{cosmology}
The most general form for the Lagrangian density with two
derivatives acting on the four-vector field $A^{\mu}$
 can be written as
\begin{eqnarray}
\label{eq:lagr}
\mathscr{L}&=&-c_1\nabla_{\mu}A^{\nu}\nabla^{\mu}A_{\nu}-c_2\left(\nabla_{\mu}A^{\mu}\right)^2
-c_3\nabla_{\mu}A^{\nu}\nabla_{\nu}A^{\mu}\nonumber\\&&+\lambda\left(\nabla_{\mu}A^{\mu}-\frac{1}{2}\gamma\lambda\right)\;.
\end{eqnarray}
Here $c_i$ are dimensionless constants. The stability analysis of
the theory and the phenomenological investigations could put
constraints on the sign and value of $c_i$.

Following the Einstein-aether theory \cite{carroll:2004}, If we
define
\begin{eqnarray}
J^{\mu}_{\ \
\nu}&\equiv&-c_1\nabla^{\mu}A_{\nu}-c_2\delta^{\mu}_{\nu}\nabla_{\alpha}A^{\alpha}
-c_3\nabla_{\nu}A^{\mu}\;,\nonumber\\
K^{\mu}_{\ \ \nu}&\equiv&\lambda\delta^{\mu}_{\nu}\;,
\end{eqnarray}
we can rewrite the Lagrangian density as follows
\begin{eqnarray}
\mathscr{L}&=&J^{\mu}_{\ \ \nu}\nabla_{\mu}A^{\nu}+K^{\mu}_{\ \
\nu}\nabla_{\mu}A^{\nu}-\frac{1}{2}\gamma\lambda^2\;.
\end{eqnarray}
Variation of the Lagrangian density with respect to $A^{\mu}$
leads to the equation of motion for $A^{\mu}$
\begin{eqnarray}\label{eq:eom}
\nabla_{\mu}J^{\mu}_{\ \ \nu}+\nabla_{\mu}K^{\mu}_{\ \
\nu}+\frac{1}{2}\nabla_{\nu}\lambda=0\;.
\end{eqnarray}
On the other hand, variation of the Lagrangian density with
respect to $\lambda$ gives the generalized Lorenz gauge condition
\begin{eqnarray}\label{eq:gauge}
\nabla_{\mu}A^{\mu}-\gamma\lambda=0\;,
\end{eqnarray}
from which we can derive the Lagrange multiplier field $\lambda$.
Acting on both sides of Eq.~(\ref{eq:gauge}) by $\nabla_{\nu}$ and
using Eq.~(\ref{eq:eom}), we find
\begin{eqnarray}\label{eq:eomA}
\gamma\nabla_{\mu}J^{\mu}_{\ \ \nu}+\gamma\nabla_{\mu}K^{\mu}_{\ \
\nu}+\frac{1}{2}\nabla_{\nu}\left(\nabla_{\mu}A^{\mu}\right)=0\;.
\end{eqnarray}
This equation determines the dynamics of $A^{\mu}$ subject to the
generalized Lorenz gauge condition.

The energy-momentum tensor for the vector field is found to be
\begin{eqnarray}\label{eq:stress}
T_{\mu\nu}&=&2c_1\left(\nabla_{\alpha}A_{\mu}\nabla^{\alpha}A_{\nu}-\nabla_{\mu}A_{\alpha}\nabla^{\mu}A_{\alpha}\right)\nonumber\\&&
+2\nabla_\sigma\left[J^{\ \ \sigma}_{(\mu}A_{\nu )}-J^{\sigma}_{\
\ (\mu}A_{\nu )}-J_{(\mu\nu)}A^{\sigma
}\right]\nonumber\\&&+\nabla_\sigma\left[K^{\ \
\sigma}_{(\mu}A_{\nu )}-K^{\sigma}_{\ \ (\mu}A_{\nu
)}-K_{(\mu\nu)}A^{\sigma }\right]\nonumber\\&&+
g_{\mu\nu}\mathscr{L}\;.
\end{eqnarray}
In the background of spatially flat Friedmann-Robertson-Walker
(FRW) Universe, the nonvanishing component of the vector is the
timelike component. So the vector field $A^{\mu}$ can be written
as
\begin{eqnarray}
A^{\mu}=\left[\phi\left(t\right),\ 0,\ 0,\ 0\right]\;.
\end{eqnarray}
Then the equation of motion, Eq.~(\ref{eq:eomA}), becomes
\begin{eqnarray}\label{eq:eomphi}
&&-\left(c_{1}+c_{2}+c_{3}-\frac{1}{2\gamma}\right)\left(\ddot{\phi}+3H\dot{\phi}+3\dot{H}\phi\right)\nonumber\\&&-3H^2\phi\left(c_1+c_3\right)=0\;,
\end{eqnarray}
where
\begin{eqnarray}
H=\frac{\dot{a}}{a}\;,
\end{eqnarray}
is the Hubble parameter. $a(t)$ is the scale factor of the
Universe. Dot denotes the derivative with respect to the cosmic
time $t$. If
\begin{eqnarray}
\label{eq:cond} c_1+c_3=0\;,
\end{eqnarray}
the equation of motion, Eq.~(\ref{eq:eomphi}), reduces to a very
simple case
\begin{eqnarray}\label{eq:eomsim}
&&\ddot{\phi}+3H\dot{\phi}+3\dot{H}\phi=0\;.
\end{eqnarray}
Solving the equation, we obtain
\begin{eqnarray}\label{eq:eomLambda}
&&\dot{\phi}+3H{\phi}=\Lambda\;,
\end{eqnarray}
where $\Lambda$ is an integration constant which has the dimension
of $l^{-1}$. Thus under the condition that Eq.~(\ref{eq:cond}) is
satisfied, the Lagrangian, Eq.~(\ref{eq:lagr}) turns out to be
\begin{eqnarray}
\label{eq:lagM}
\mathscr{L}&=&-\frac{c_1}{2}F_{\mu\nu}F^{\mu\nu}-c_2\left(\nabla_{\mu}A^{\mu}\right)^2
\nonumber\\&&+\lambda\left(\nabla_{\mu}A^{\mu}-\frac{1}{2}\gamma\lambda\right)\;,
\end{eqnarray}
where
\begin{eqnarray}
F_{\mu\nu}&=&\nabla_{\mu}A_{\nu}-\nabla_{\nu}A_{\mu}\;,
\end{eqnarray}
is the field strength tensor.

In this paper, we are interested in the case of $c_1+c_3=0$ for
simplicity. From the generalized Lorenz gauge condition,
Eq.~(\ref{eq:gauge}), we obtain
\begin{eqnarray}\label{eq:GLG}
&&\dot{\phi}+3H{\phi}=\gamma\lambda\;.
\end{eqnarray}
Taking account of Eq.~(\ref{eq:eomLambda}), we find the Lagrange
multiplier field is actually a constant,
\begin{eqnarray}
\lambda=\frac{\Lambda}{\gamma}\;.
\end{eqnarray}
So we can solve for $\phi$ from Eq.~(\ref{eq:GLG}) as follows
\begin{eqnarray}
\phi=\frac{\Lambda}{a^3}\int a^3dt\;.
\end{eqnarray}

Then the energy density and pressure of the vector field derived
from the energy-momentum tensor take the form
\begin{eqnarray}
\label{eq:CC}
\rho_{A}&=&-T_{0}^{0}=\left(-c_2+\frac{1}{2\gamma}\right)\Lambda^2\;,\nonumber\\
p_{A}&=&T_{i}^{i}=-\left(-c_2+\frac{1}{2\gamma}\right)\Lambda^2\;.
\end{eqnarray}
This is exactly for the energy density and pressure of Einstein's
cosmological constant. To interpret the current acceleration of
the Universe, we expect $\Lambda$ to be the order of the inverse
of present-day Hubble length and

\begin{eqnarray}
-c_2+\frac{1}{2\gamma}>0\;.
\end{eqnarray}
If $c_2=0$, we conclude that $\gamma$ must be positive. The
Feynman and Yennie-Fried gauge satisfy this requirement, while the
Landau gauge leads to a vanishing cosmological constant.
\subsection{static and spherically symmetric solution}
In order to show the four-vector field theory can pass the solar
system tests, in this subsection, let's seek for the static and
spherically symmetric solution of Einstein equations sourced by
the Lagrangian Eq.~(\ref{eq:lagM}). The static and spherically
symmetric metric can always be written as
\begin{eqnarray}
ds^2=-U\left(r\right)dt^2+\frac{1}{U\left(r\right)}dr^2+f\left(r\right)^2d\Omega^2\;.
\end{eqnarray}
Comparing to solving the Einstein equations, we prefer to start
from the Lagrangian Eq.~(\ref{eq:lagM}) for simplicity in
calculations. Because of the static and spherically symmetric
property of the spacetime, the vector field $A_{\mu}$ takes the
form
\begin{eqnarray}
A_{\mu}=\left[\phi\left(r\right),\ \psi\left(r\right),\ 0,\
0\right]\;,
\end{eqnarray}
where $\phi$ and $\psi$ correspond to the electric and magnetic
part of the electromagnetic potential.  Then we have
\begin{eqnarray}
F_{\mu\nu}F^{\mu\nu}=-2\phi^{'2}\;,\ \ \
\nabla_{\mu}A^{\mu}=\left(U\psi\right)^{'}+2\frac{f^{'}}{f}U\psi\;,
\end{eqnarray}
where prime denotes the derivative with respect to $r$. Taking
into account the Ricci scalar, $R$, we have the total Lagrangian
from Eq.~(\ref{eq:lagM})

\begin{eqnarray}
\label{eq:lagMM}
\mathscr{L}&=&-U^{''}-4U^{'}\frac{f^{'}}{f}-4U\frac{f^{''}}{f}+\frac{2}{f^2}-2U\frac{f^{'2}}{f^2}
\nonumber\\&&+{c_1}\phi^{'2}-c_2\left[\left(U\psi\right)^{'}+2\frac{f^{'}}{f}U\psi\right]^2
\nonumber\\&&+\lambda\left[\left(U\psi\right)^{'}+2\frac{f^{'}}{f}U\psi-\frac{1}{2}\gamma\lambda\right]\;.
\end{eqnarray}
Using the Euler-Lagrange equation, we obtain the equation of
motion for $\phi$,

\begin{eqnarray}
\label{eq:phi} f\phi^{''}+2\phi^{'}f^{'}=0\;,
\end{eqnarray}
for $\lambda$,

\begin{eqnarray}\label{eq:lambda}
\left(U\psi\right)^{'}+2\frac{f^{'}}{f}U\psi-\gamma\lambda=0\;,
\end{eqnarray}
for $U$,

\begin{eqnarray}\label{eq:U}
&&2ff^{''}-4c_2f\psi^2U^{'}f^{'}-4c_2\psi
f^2U^{'}\psi^{'}-2c_2\psi^2f^2U^{''}\nonumber\\&&-4c_2\psi U f
\psi^{'}f^{'}+4c_2U\psi^2f^{'2}-2c_2\psi U
f^2\psi^{''}\nonumber\\&&-4c_2Uf\psi^{2}f^{''}+\psi
f^2\lambda^{'}=0\;,
\end{eqnarray}

for $\psi$,

\begin{eqnarray}\label{eq:psi}
&&-4c_2U\psi ff^{''}-4c_2f\psi U^{'}f^{'}-4c_2
f^2U^{'}\psi^{'}-2c_2\psi f^2U^{''}\nonumber\\&&
 -4c_2 U f\psi^{'}f^{'}+4c_2U\psi f^{'2}-2c_2 U f^2\psi^{''}\nonumber\\&&+
f^2\lambda^{'}=0\;,
\end{eqnarray}

and for$f$,
\begin{eqnarray}\label{eq:f}
&&-2c_1f\phi^{'2}-4c_2Uf\psi^2U^{''}-12c_2U\psi f
U^{'}\psi^{'}-4c_2U^2\psi
f\psi^{''}\nonumber\\&&-16c_2U\psi^2U^{'}f^{'}-16c_2U^2\psi\psi^{'}f^{'}-8c_2U^2\psi^2f^{''}
\nonumber\\&&-2c_2f\psi^2U^{'2}+\gamma f
\lambda^2+4U^{'}f^{'}+4Uf^{''}+2fU^{''}\nonumber\\&&+2Uf\psi\lambda^{'}-2c_2fU^2\psi^{'2}\;,
\end{eqnarray}
respectively. Now we have five independent differential equations
and five unknown variables, $U,\ f,\ \phi,\ \psi,\ \lambda$. So
the system of equations is closed.

From Eq.~(\ref{eq:phi}), we obtain
\begin{eqnarray}\label{eq:phi1}
\phi=\phi_1+\phi_0\int\frac{1}{f^2}dr\;,
\end{eqnarray}
where $\phi_0,\ \phi_1$ are two integration constants. In order to
fix $\phi_1$, let us consider the asymptotic condition, namely,
for large enough $r$. We expect to have $f^2\sim r^2$. So
$\phi\sim \phi_1+\phi_0/r$. Let the asymptotic value of $\phi=0$
when $r$ is large enough. Then we conclude that

\begin{eqnarray}\label{eq:phi2}
\phi_1=0\;.
\end{eqnarray}

On the other hand, from Eq.~(\ref{eq:lambda}), we obtain

\begin{eqnarray}\label{eq:lambda2}
\lambda=\frac{1}{\gamma}\left[\left(U\psi\right)^{'}+2\frac{f^{'}}{f}U\psi\right]\;.
\end{eqnarray}
Keeping Eqs.~(\ref{eq:phi1}), (\ref{eq:phi2}) and
(\ref{eq:lambda2}) in mind, we obtain from the difference of
Eq.(\ref{eq:U}) and Eq.~(\ref{eq:psi})

\begin{eqnarray}
f^{''}=0\;.
\end{eqnarray}

So we have

\begin{eqnarray}
f=f_0+f_1r\;.
\end{eqnarray}
$f_0,\ f_1$ are two integration constants. We can always rescale
$r$ such that

\begin{eqnarray}
f_0=0\;, \ \ f_1=1\;.
\end{eqnarray}

Then we obtain from Eq.~(\ref{eq:psi})

\begin{eqnarray}
\psi=\frac{\psi_0r}{U}+\frac{\psi_1}{r^2U}\;,
\end{eqnarray}
with $\psi_0,\ \psi_1$ two integration constants. Finally,
Eq.~(\ref{eq:f}) turns out to be
\begin{eqnarray}
&&2\gamma r^4 U^{''}+4\gamma
r^3U^{'}+9\psi_0^2r^4-2c_1\gamma\phi_0^2\nonumber\\&&-18c_2\gamma
\psi_0^2 r^4=0\;,
\end{eqnarray}
from which we obtain

\begin{eqnarray}
U=U_0-\frac{U_1}{r}+\frac{c_1\phi_0^2}{2r^2}-\frac{3}{2}
\left(-c_2+\frac{1}{2\gamma}\right)\psi_0^2r^2\;,
\end{eqnarray}
where $U_0$ and $U_1$ are two integration constants. Compare it
with the Reissner-Nordstrom-de Sitter solution
\begin{eqnarray}
U=1-\frac{2M}{r}+\frac{Q^2}{r^2}-\frac{1}{6}Lr^2\;,
\end{eqnarray}
we may put

\begin{eqnarray}
&&U_0=1\;, \ \ \ U_1=2M\;,\ \ \ c_1=2\;,\ \ \phi_0=Q\;,\ \
\nonumber\\&& \psi_0=\frac{\Lambda}{3}\;,\ \ \
L=\left(-c_2+\frac{1}{2\gamma}\right)\Lambda^2\;.
\end{eqnarray}
Here $L$ denotes the cosmological constant.

Therefore, the static and spherically solution for the Lagrangian
density, Eq.~(\ref{eq:lagM}), is exactly the Reissner-Nordstrom-de
Sitter solution. The field $\phi,\ \psi$ and the Lagrange
multiplier $\lambda$ is found to be
\begin{eqnarray}
&&\phi=\frac{Q}{r}\;, \ \ \ \lambda=\frac{\Lambda}{\gamma}\;,\ \
\psi=\frac{\Lambda r}{3U}+\frac{\psi_1}{r^2U}\;.
\end{eqnarray}
We recognize that $\phi$ is the electric potential sourced by the
change $Q$. The cosmological constant is closely related to the
field strength $\partial_r(r^2U \psi)$ of the magnetic potential
$\psi$. Since the static and spherically symmetric solution of the
theory is just the Reissner-Nordstrom-de Sitter solution, we
conclude that it would not violate the solar system tests on
gravity theory.
\section{nonminimally coupled to gravitation}
\subsection{equations of motion and energy momentum tensor} When the four-vector $A^{\mu}$ is
nonminimally coupled to the gravitation, we have the Lagrangian
density as follows:

\begin{eqnarray}
\label{eq:lagC}
\mathscr{L}&=&-\frac{c_1}{2}F_{\mu\nu}F^{\mu\nu}-c_2\left(\nabla_{\mu}A^{\mu}\right)^2+b_1
R_{\mu\nu}A^{\mu}A^{\nu} \nonumber\\&&+b_2 R
A_{\mu}A^{\nu}+\lambda\left(\nabla_{\mu}A^{\mu}-\frac{1}{2}\gamma\lambda\right)\;.
\end{eqnarray}
Here $b_1,\ b_2$ are two dimensionless constants and $R_{\mu\nu},\
R$ are the Ricci tensor and the Ricci scalar, respectively. In the
first place, varying the Lagrangian density with respect to
$A^{\mu}$, we obtain the equation of motion for $A^{\mu}$

\begin{eqnarray}
\label{eq:eomC1} && c_1\nabla_{\nu} F^{\nu}_{\ \
\mu}+c_2\nabla_{\mu}\left(\nabla_{\nu}A^{\nu}\right)+b_1
R_{\mu\nu}A^{\nu} \nonumber\\&&+b_2 R
A_{\mu}-\frac{1}{2}\nabla_{\mu}\lambda=0\;.
\end{eqnarray}
Secondly, varying the Lagrangian density with respect to
$\lambda$, we obtain the equation of motion for $\lambda$

\begin{eqnarray}
\label{eq:eomC2} \nabla_{\mu}A^{\mu}-\gamma\lambda=0\;.
\end{eqnarray}
Finally, varying the Lagrangian density with respect to
$g^{\mu\nu}$, we obtain the energy momentum

\begin{eqnarray}\label{eq:stressC}
T_{\mu\nu}&=&2\nabla_\sigma\left[J^{\ \ \sigma}_{(\mu}A_{\nu
)}-J^{\sigma}_{\ \ (\mu}A_{\nu )}-J_{(\mu\nu)}A^{\sigma
}\right]\nonumber\\&&+\nabla_\sigma\left[K^{\ \
\sigma}_{(\mu}A_{\nu )}-K^{\sigma}_{\ \ (\mu}A_{\nu
)}-K_{(\mu\nu)}A^{\sigma
}\right]\nonumber\\&&+b_1\left[2\nabla_{\sigma}\nabla_{(\mu}
\left(A_{\nu {)}}A^{\sigma}\right)-\nabla_{\alpha}\nabla_{\sigma}
\left(A^{\alpha}A^{\sigma}\right)g_{\mu\nu}\right.\nonumber\\
&&\nonumber \\
&&\left.-\nabla^{2}\left(A_{\mu}A_{\nu}\right)-4A^{\sigma}R_{\sigma(\mu}
A_{\nu)}\right]-2b_2\left[R_{\mu\nu}A_{\sigma}A^{\sigma}\right.\nonumber\\
&&\nonumber \\
&&\left.+RA_{\mu}A_{\nu}-\nabla_{\mu}\nabla_{\nu}\left(A_{\sigma}A^{\sigma}\right)+\nabla^2\left(A_{\sigma}A^{\sigma}\right)g_{\mu\nu}\right]\nonumber\\&&+2c_1F_{\mu\sigma}F^{\sigma}_{\
\ \nu}+ g_{\mu\nu}\mathscr{L}\;.
\end{eqnarray}
Here $J_{\mu\nu}$ is understood with $c_1+c_3=0$. Combining
Eq.~(\ref{eq:eomC1}) and Eq.~(\ref{eq:eomC2}), we can express the
equation of motion for $A^{\mu}$ as follows
\begin{eqnarray}
\label{eq:eomCC} && c_1\nabla_{\nu} F^{\nu}_{\ \
\mu}+c_2\nabla_{\mu}\left(\nabla_{\nu}A^{\nu}\right)+b_1
R_{\mu\nu}A^{\nu} \nonumber\\&&+b_2 R
A_{\mu}-\frac{1}{2\gamma}\nabla_{\mu}\left(\nabla_{\nu}A^{\nu}\right)=0\;.
\end{eqnarray}

\subsection{cosmic evolution}

In the background of spatially flat FRW Universe, the equations of
motion, Eq.~(\ref{eq:eomCC}) takes the form
\begin{eqnarray}
\label{eq:eomC3}
&&\left(c_2-\frac{1}{2\gamma}\right)\left(\dot{\phi}+3H\phi\right)^{\cdot}-3\left(b_1+2b_2\right)\dot{H}\phi\nonumber\\&&-3H^2\phi\left(b_1+4b_2\right)=0\;.
\end{eqnarray}

From the energy momentum tensor, we obtain the energy density and
pressure of the four-vector field $A^{\mu}$

\begin{eqnarray}\label{eq:denC}
\rho_{A}&=&c_2\left(2\phi\ddot{\phi}-\dot{\phi}^2\right)+6\left(b_1+2b_2\right)H\phi\dot{\phi}\nonumber\\&&
+\left[6\left(c_2-b_1-2b_2\right)\dot{H}-9\left(c_2+2b_2\right)H^2\right]\phi^2\nonumber\\&&-\phi\dot{\lambda}+\frac{1}{2}\gamma\lambda^2\;,
\end{eqnarray}

\begin{eqnarray}\label{eq:prC}
p_{A}&=&2\left(c_2-b_1-2b_2\right)\left(\phi\ddot{\phi}+\dot{\phi}^2\right)-c_2\dot{\phi}^2\nonumber\\&&+\left(3c_2-2b_1-2b_2\right)\left[4H\phi\dot{\phi}
+\left(2\dot{H}+3H^2\right)\phi^2\right]\nonumber\\&&-\phi\dot{\lambda}+\frac{1}{2}\gamma\lambda^2\;.
\end{eqnarray}

Using the equation of motion for $A^{\mu}$, Eq.~(\ref{eq:eomC3}),
and the generalized Lorentz gauge condition, Eq.~(\ref{eq:eomC2}),
we can eliminate $\ddot{\phi}$ and rewrite the energy density as
following

\begin{eqnarray}\label{eq:denD}
\rho_{A}&=&6H\dot{\phi}\phi\left(-c_2
+\frac{1}{2\gamma}+b_1+2b_2\right)\nonumber\\&&+9H^2\phi^2\left(-c_2+
\frac{1}{2\gamma}+\frac{2}{3}b_1+\frac{2}{3}b_2\right)\nonumber\\&&
+\left(-c_2+\frac{1}{2\gamma}\right)\dot{\phi}^2\;.
\end{eqnarray}
We have verified that the energy density, Eq.~(\ref{eq:denD}), and
the pressure, Eq.~(\ref{eq:prC}), satisfy the energy conservation
equation

\begin{eqnarray}\label{eq:ece}
\frac{d\rho_{A}}{dt}+3H\left(\rho_A+p_A\right)=0\;,
\end{eqnarray}
which is consistent with the equation of motion,
Eq.~(\ref{eq:eomC3}) and the energy momentum conservation equation

\begin{eqnarray}\label{eq:emc}
\nabla_{\nu}T^{\nu}_{\ \ \mu}=0\;.
\end{eqnarray}
In other words, the equation of motion Eq.~(\ref{eq:eomC3}), the
energy conservation equation Eq.~(\ref{eq:ece}) and the energy
momentum conservation equation Eq.~(\ref{eq:emc}) give actually
the same equation. Similarly, using the equation of motion for
$A^{\mu}$, Eq.~(\ref{eq:eomC3}), and the generalized Lorentz gauge
condition, Eq.~(\ref{eq:eomC2}), we can eliminate $\ddot{\phi}$
and rewrite the pressure as following

\begin{eqnarray}\label{eq:prD}
p_{A}&=&-\left(-c_2+\frac{1}{2\gamma}+2b_1+4b_2\right)\dot{\phi}^2\nonumber\\&&
+\left(6c_2+4b_2-\frac{3}{\gamma}-2b_1\right)H\phi\dot{\phi}\nonumber\\&&
-\frac{4\gamma}{2c_2\gamma-{1}}\left(\frac{2b_1}{\gamma}+\frac{5b_2}{\gamma}+12b_2^2+12b_1b_2\right.\nonumber\\
&&\nonumber \\
&&\left.+3b_1^2-10c_2b_2-4c_2b_1
\right)\dot{H}\phi^2\nonumber\\&&-\frac{3\gamma}{4c_2\gamma-2}H^2\phi^2\left(-\frac{3}{\gamma^2}+\frac{12b_2+12c_2}{\gamma}\right.\nonumber\\
&&\nonumber \\
&&\left.+64b_2^2-12c_2^2+48b_1b_2+8b_1^2-24c_2b_2\right)\;.
\end{eqnarray}

Then we find

\begin{eqnarray}\label{eq:state}
&&p_{A}-\frac{{1}-\left(2c_2-4b_1-8b_2\right)\gamma}{2c_2\gamma-1}\rho_A\nonumber\\&=&\frac{4}{1-2c_2\gamma}\phi\left(\phi\dot{H}+2H\dot{\phi}+3H^2\phi\right)\nonumber\\&&
\cdot\left[\left(3b_1^2+12b_2^2-10c_2b_2+12b_1b_2-4c_2b_1\right)\gamma\right.\nonumber\\
&&\nonumber \\
&&\left.+2b_1+5b_2\right]\;.
\end{eqnarray}

It is apparent if
\begin{eqnarray}
b_1=b_2=0\;,
\end{eqnarray}
we have the equation of state for the four-vector $A^{\mu}$
\begin{eqnarray}
\omega_{A}=\frac{p_{A}}{\rho_{A}}=-1\;,
\end{eqnarray}
which is consistent with the result of Eq.~(\ref{eq:CC}).

On the other hand, if $b_1\neq 0$ (or $b_2\neq 0$) and
\begin{eqnarray}
&&\left(3b_1^2+12b_2^2-10c_2b_2+12b_1b_2-4c_2b_1\right)\gamma\nonumber\\&&+2b_1+5b_2=0\;,
\end{eqnarray}
namely,

\begin{eqnarray}
\label{eq:gamma99}
\gamma=-\frac{2b_1+5b_2}{3b_1^2+12b_2^2-10c_2b_2+12b_1b_2-4c_2b_1}\;,
\end{eqnarray}

we have the equation of state

\begin{eqnarray}
\omega_{A}&=&\frac{5b_1+14b_2}{3\left(b_1+2b_2\right)}\nonumber\\
&=&\frac{14+5\frac{b_1}{b_2}}{3\left(2+\frac{b_1}{b_2}\right)}\nonumber\\
&=&\frac{14+5r}{3\left(2+r\right)}\;,
\end{eqnarray}
where we define the ratio $r$ of $b_1$ and $b_2$ as $r=b_1/b_2$.
In other words, if the value of $\gamma$ is given by
Eq.~(\ref{eq:gamma99}), we shall have a constant equation of state
for vector field $A^{\mu}$.

\begin{figure}
\includegraphics[width=8.4cm]{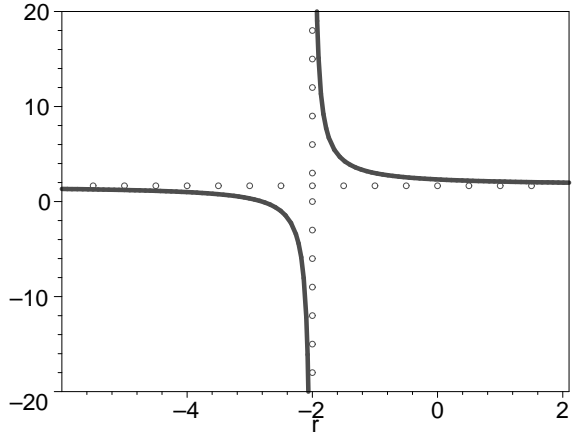}
\\
\caption{The equation of state $w_A$ of the four-vector field
$A^{\mu}$ with the ratio $r$ of $b_1$ and $b_2$. It is not bounded
both below and up, namely, $-\infty<w_A<+\infty$. } \label{fig:wh}
\end{figure}

In Fig.~\ref{fig:wh}, we plot the equation of state $\omega_{A}$
of the four-vector field $A^{\mu}$ with the ratio $r$ of $b_1$ and
$b_2$. It is not bounded both below and up, namely,
$-\infty<w<+\infty$. This is different from the quintessence which
has the equation of state $-1\leq w \leq+1$.

In particular, when

\begin{eqnarray}
b_2=-\frac{5}{14}b_1\;,\ \ \
\gamma=\frac{7}{2\left(7c_2-4b_1\right)}\;,
\end{eqnarray}

we have the equation of state

\begin{eqnarray}
\omega_A=0\;,
\end{eqnarray}
which indicates that the four-vector field $A^{\mu}$ behaves as a
dust matter. So the four-vector field could behave as vast matters
in the background of spatially flat FRW Universe. However, this
may be not the case. The weak, strong and dominant energy
conditions \cite{wald:84}, the stability analysis of classical and
quantum aspects would put constraints on the parameters and so the
equation of state would be greatly constrained.

\section{Conclusion and discussion}
In conclusion, motivated by the Lagrange multiplier method in the
Einstein-aether theory, we introduce the generalized Lorentz gauge
condition in the theory of quantum electrodynamics into the
general vector-tensor theories of gravity. With the fix-norm
constraint, it is found that the Einstein-aether field contributes
an energy density \cite{carroll:2004}
\begin{eqnarray}
\rho_A=-3\left(c_1+3c_2+c_3\right)m^2H^2\;,
\end{eqnarray}
where the constant $m$ represents the norm of the four-vector
$A^{\mu}$. In order that the energy density be positive, one
should have $c_1+3c_2+c_3<0$. However, as shown by Lim, the
investigations of quantum aspect on this theory require
$c_1+3c_2+c_3\geq 0$ \cite{lim:04}. Therefore, the energy density
of the aether field is nonpositive which violets the weak energy
condition \cite{wald:84}.

However, with the generalized Lorentz gauge condition, we find
that the vector filed could contribute a constant energy density.
For the minimally coupled case, in particular, when $c_1=-c_3,\
c_2=0$, we always have a non-negative energy density for the
Landau gauge, Feynman gauge and Yennie-Fried gauge. The
corresponding quantum aspects have been very well studied. On the
other hand, the static and spherically symmetric solution sourced
by the vector field is exactly the Reissner-Nordstrom-de Sitter
solution. This reveals that the theory may not be conflict with
the solar system tests on gravity theory. For the non-minimally
coupled case, the vector may play the role of vast matters in the
background of spatially flat Universe. But it may be not the case.
The stability analysis of classical and quantum aspects would
surely constrain the space of parameters. It is a very important
issue. We plan to leave this analysis for future publications.

 \acknowledgments

We thank  A. L. Maroto and J. B. Jimenez for helpful discussions.
This work is supported by the National Science Foundation of China
under the Key Project Grant No. 10533010, Grant No. 10575004,
Grant No. 10973014, and the 973 Project (No. 2010CB833004).

\newcommand\ARNPS[3]{~Ann. Rev. Nucl. Part. Sci.{\bf ~#1}, #2~ (#3)}
\newcommand\AL[3]{~Astron. Lett.{\bf ~#1}, #2~ (#3)}
\newcommand\AP[3]{~Astropart. Phys.{\bf ~#1}, #2~ (#3)}
\newcommand\AJ[3]{~Astron. J.{\bf ~#1}, #2~(#3)}
\newcommand\APJ[3]{~Astrophys. J.{\bf ~#1}, #2~ (#3)}
\newcommand\APJL[3]{~Astrophys. J. Lett. {\bf ~#1}, L#2~(#3)}
\newcommand\APJS[3]{~Astrophys. J. Suppl. Ser.{\bf ~#1}, #2~(#3)}
\newcommand\JHEP[3]{~JHEP.{\bf ~#1}, #2~(#3)}
\newcommand\JCAP[3]{~JCAP. {\bf ~#1}, #2~ (#3)}
\newcommand\LRR[3]{~Living Rev. Relativity. {\bf ~#1}, #2~ (#3)}
\newcommand\MNRAS[3]{~Mon. Not. R. Astron. Soc.{\bf ~#1}, #2~(#3)}
\newcommand\MNRASL[3]{~Mon. Not. R. Astron. Soc.{\bf ~#1}, L#2~(#3)}
\newcommand\NPB[3]{~Nucl. Phys. B{\bf ~#1}, #2~(#3)}
\newcommand\CQG[3]{~Class. Quant. Grav.{\bf ~#1}, #2~(#3)}
\newcommand\PLB[3]{~Phys. Lett. B{\bf ~#1}, #2~(#3)}
\newcommand\PRL[3]{~Phys. Rev. Lett.{\bf ~#1}, #2~(#3)}
\newcommand\PR[3]{~Phys. Rep.{\bf ~#1}, #2~(#3)}
\newcommand\PRD[3]{~Phys. Rev. D{\bf ~#1}, #2~(#3)}
\newcommand\RMP[3]{~Rev. Mod. Phys.{\bf ~#1}, #2~(#3)}
\newcommand\SJNP[3]{~Sov. J. Nucl. Phys.{\bf ~#1}, #2~(#3)}
\newcommand\ZPC[3]{~Z. Phys. C{\bf ~#1}, #2~(#3)}
 \newcommand\IJGMP[3]{~Int. J. Geom. Meth. Mod. Phys.{\bf ~#1}, #2~(#3)}
  \newcommand\GRG[3]{~Gen. Rel. Grav.{\bf ~#1}, #2~(#3)}

\end{document}